\documentclass[twocolumn, notitlepage,aps,showpacs,floats,amssymb,amsmath,floatfix,groupedaddress,superscriptaddress,aps,pre]{revtex4-2}
\usepackage{amsfonts,amssymb,stmaryrd,latexsym,amsmath,braket}
\usepackage{graphicx} % Required for inserting images
\usepackage{booktabs}
\usepackage{comment}
\usepackage{newtxtext, newtxmath} % Modern replacement for times font package
\usepackage{slashed}
\usepackage{bm}
\usepackage{appendix}
\usepackage{enumitem}
\usepackage{multirow}
\usepackage{array}
\usepackage{placeins}
\usepackage[dvipsnames]{xcolor}

\usepackage[colorlinks=true,linktocpage=true,citecolor=blue,urlcolor=blue,linkcolor=blue]{hyperref}
\begin{document}

\title{Demixing in a Binary System with Differential Diffusivity in Presence of External Potential: Dynamical Anisotropy and non-Gaussian Position Fluctuations }

\title{Emergence of Dynamical Anisotropy induced by Demixing in a Binary System with Differential Diffusivity under an External Potential}
\author{Rashmi Trivedi}
\affiliation{Department of Physics, Indian Institute of Technology, Jodhpur 342037, India}

\author{Subhajit Paul}\email{spaul@physics.du.ac.in}
\affiliation{Department of Physics and Astrophysics, University of Delhi, Delhi- 110007, India}

\author{Sumanta Kundu}\email{sumanta.kundu@na.infn.it}
 \affiliation{Dipartimento di Fisica, Università di Napoli Federico II, and INFN Napoli, Complesso Universitario di Monte
Sant’Angelo, 80126 Napoli, Italy}

\author{Sunita Kumari}
\email{sunita@iitj.ac.in}
\affiliation{Department of Physics, Indian Institute of Technology, Jodhpur 342037, India}

\date{\today}

%%%%%%%%%%%%%%%%%%%%%%%%%%%%%%%%%%%%%%%%%%%%%%%%%%%%%%%%%%%%%%
%%% Abstarct
%%%%%%%%%%%%%%%%%%%%%%%%%%%%%%%%%%%%%%%%%%%%%%%%%%%%%%%%%%%%%%
\begin{abstract}
Spontaneous demixing in active matter is a ubiquitous phenomenon that is crucial for numerous living processes ranging from bacterial swarming to sorting of cells in dense tissues. Here, we systematically investigate the effect of spatially varying potential acting along one direction and packing fraction on the binary mixture of particles with different diffusivities. Our results indicate that the presence of an external potential promotes demixing over a larger range of packing fractions, while also fostering a more pronounced `hexatic order' within the bands of less diffusive (``cold'') particles formed near the minima of the potential. The mean-square displacements (MSD) of ``cold" and ``hot" particles in different directions exhibit a distinct behavior. In contrast to the long-time sub-diffusive behavior of the ``cold'' particles, the ``hot'' ones display diffusive nature following an
intermediate plateau. However, in the direction transverse to the applied potential, both types of particles undergo normal diffusion. Furthermore, interesting non-Gaussian characteristics are observed, corresponding to the spatial distribution of the displacement of ``hot'' and ``cold'' particles. Interestingly, our results reveal the formation of a 'percolating band', and the emergence of such dynamic anisotropy is not observed in the absence of an external potential. These aspects are highly relevant to the dynamics of various systems—including densely packed tissues, bacterial motility in confined spaces, and granular segregation in the pharmaceutical industry.
\end{abstract}

\maketitle

%%%%%%%%%%%%%%%%%%%%%%%%%%%%%%%%%%%%%%%%%%%%%%%%%%%%%%%%%%%%%%%%%%%%%%%%%%%%%%%%%%%%%%%%%%%%%%%%%%%%%%%%%%%%%%%%%%%%%%%%%%%%%%%%%%%%%%%%%%%%%
\section{Introduction}
%%%%%%%%%%%%%%%%%%%%%%%%%%%%%%%%%%%%%%%%%%%%%%%%%%%%%%%%%%%%%%%%%%%%%%%%%%%%%%%%%%%%%%%%%%%%%%%%%%%%%%%%%%%%%%%%%%%%%%%%%%%%%%%%%%%%%%%%%%%%
Demixing phenomena or phase-separation in active matter systems appears quite naturally and plays a significant role in understanding the spatial organizations in various biological contexts~\cite{bechinger2016active,vicsek_physrep_12,Romanczuk2012,Cates2015,Vicsek1995,peruani2006_pre,fily_12,barberis2019phase,Paul2021,caprini2023flocking,gaur2026global}. It models how cells orchestrate biochemistry through biomolecular condensates, tissue compartmentalization, and how populations of proliferating active matter (such as bacteria or cancer cells) organize, segregate, and spread~\cite{bechinger2016active}.  However, similar demixing phenomena are also observed in many other scenarios such as clustering and segregation of granular particles~\cite{brilliantov_prl_2004,Melby2005,Rivas_2011,paul2014dynamics}, self-assembly of colloids~\cite{Mao1995,Biben1996,Dijkstra1994}, biological reaction-diffusion processes~\cite{Wheeler2018,Su2016,Brangwynne2009,Finck2016}, separation in colloid-polymer mixtures~\cite{Zhang2013, Poon2002}, emergence of dense aggregates along a polymer chain or protein during its collapse~\cite{guo2011coil,reddy2017collapse,paul2022effects}, coarsening during phase separation in a binary system of spins or in a fluid ~\cite{das2006molecular,das2012finite,das2020initial}, etc. Even though the large-scale patterns may look similar, the microscopic dynamics and nonequilibrium pathways differ across different systems \cite{elgeti2015physics,bechinger2016active,gollub_rmp_1999}.  For example, external vibration helps demix granular particles of different shapes and sizes ~\cite{Liu2022, Kudrolli2004, Menbari2020}. On the other hand, an assembly of active particles that interact purely repulsively shows motility-induced phase separation (MIPS) \cite{Cates2015,mcand_12,mishra_12,redner2013structure,levis2018micro,mandal2019motility,paul2024spontaneous}. This process occurs spontaneously due to persistent self-propulsion, thereby establishing a feedback mechanism between slowing down and increasing crowding. 

In presence of interactions, whether these active agents forms a motile cluster or an elongated polymer chain, their dynamics as well as the tagged particle behavior, such as, anomalous diffusion, non-Gaussian position and velocity fluctuations, etc. appear very intriguing and require framework of non-equilibrium physics \cite{Vicsek1995,chate_08,Romanczuk2012, Cates2015,bechinger2016active,Br2020,Chat2020,fily_12,  Speck2014, Siebert2017,Caprini2020, paul2024dynamical,khali_pre2024,Knippenberg2024,majumder2024enhanced,Jhajhria2025,patel2026crossover}. Even though  at long times, the behavior of an active particle becomes largely indistinguishable from that the passive particles, and the effect of the self-propulsion force can be mapped to an enhancement of diffusivity. Unlike the diffusion of a passive particle, the self-propelled persistence motion of an active particle shows a short-time ballistic behavior \cite{paul2024dynamical,Romanczuk2012}. The self-organization and glassy dynamics of cells are modeled as a system of dense active particles~\cite{berthier2019glassy,henkes2020dense}. In processes such as embryonic development, tumor formation, and wound healing, cells and tissues undergo an `epithelial-to-mesenchymal transition'; that is, they transform from a more rigid, solid-like state into a more fluid state \cite{Pasupalak2020}. This naturally raises the question of whether an ensemble of particles identical in size, morphology, mass, and interactions can exhibit spontaneous demixing; what parameters govern this behavior; and how an external potential influences their dynamics.

A binary mixture of active particles with different propulsion strengths can undergo demixing and self-organize at intermediate and high densities \cite{dolai2018phase}. Interestingly, without any attractive forces identical passive Brownian particles can exhibit similar phase separation, i.e., ``cold'' clusters surrounded by a ``hot'' gas phase made of particles with higher diffusivity \cite{weber2016demix, kumari2017demixing,ilker2020phase,McCarthy2024}. Here, demixing occurs solely due to the difference in diffusivities if one of the species has much higher diffusivity compared to the other. This represents a fundamentally nonequilibrium process: even in the absence of any persistent motion of the passive particles, the unequal coupling to the thermal bath breaks detailed balance and generates effective diffusive fluxes that drive phase separation.

Considering the possibility of mapping the strength of self-propulsion in terms of long-time effective diffusivity \cite{loi_effect_2008}, we consider a binary system of Brownian particles rather than considering self-propelling particles. These are identical in shape and size but differ only in their diffusion coefficients. This system is subjected to an external potential that can modulate the spatial distribution of particles and has an effect on the demixing properties. The ``hot'' particles with higher diffusivity explore the landscape more rapidly than the ``cold'' ones, leading to the possibility of clustering. The presence of  an external potential can induce a potential-assisted accumulation for barrier-induced segregation~\cite{kumari2017demixing}. By systematically varying the diffusivity ratio and the packing fraction over a wide range, we aim to understand how the dynamics and correspondingly the demixing properties change in this externally structured environment. Interestingly, with a higher packing fraction and potential, the cluster of ``cold'' particles tends to form a percolating band in the transverse direction of the applied potential. 

Along with anomalous diffusion of ``cold'' particles which are within the cluster, the motion of ``hot'' particles also gets substantially affected. Whereas due to pressure gradient the ``cold'' particles stays near the minima of the potential, the percolating band, on the other hand, works as a reflecting barrier to the ``hot'' particles \cite{kumari2017demixing}. This creates dynamical anisotropy in the motion of ``cold'' and ``hot'' particles, and the corresponding mean-squared-displacement (MSD) becomes different in different directions. In terms of full displacement distribution,  various intermediate non-Gaussian features are observed compared to  Gaussian behavior in the case of free diffusion. Inside the cluster, nice hexagonal arrangement of ``cold'' particles is observed.  Our results can be applied and also contribute to the understanding of nonequilibrium pathways for demixing mechanisms, and shed light on analogous processes in biological, granular, and synthetic soft-matter systems, where components can exhibit different mobilities under identical external conditions.

This paper is organized as follows.  The model and the simulation details are presented in Sec.~II. The results are discussed in detail Sec.~III. Finally, in Sec.~IV we summarize our results with a few possible future prospects.  

%%%%%%%%%%%%%%%%%%%%%%%%%%%%%%%%%%%%%%%%%%%%%%%%%%%%%%%%%%%%%%%%%%%%%%%%%%%%%%%%%%%%%%%%%%%%%%%%%%%%%%%%%%%%%%%%%%%%%%%%%%%%%%%%%%%%%%%%%%%%%%%
\section{Model and Simulation}
\label{sec:Model}
%%%%%%%%%%%%%%%%%%%%%%%%%%%%%%%%%%%%%%%%%%%%%%%%%%%%%%%%%%%%%%%%%%%%%%%%%%%%%%%%%%%%%%%%%%%%%%%%%%%%%%%%%%%%%%%%%%%%%%%%%%%%%%%%%%%%%%%%%%%%%%%%%
We consider a binary mixture of particles characterized by two distinct diffusivities in a two-dimensional box of size $L \times L$ with periodic boundary conditions in both directions. The system contains $N_c$ ``cold'' particles (with a lower diffusion coefficient, $D_{\text{cold}}$) and $N_h$ ``hot''  particles (with a higher  diffusion coefficient, $D_{\text{hot}}$), so that the total number of particles $N=N_c+N_h$. We have chosen  $N_c = N_h = N/2$. We fix $D_{\text{hot}}=5$ to ensure liquid-like dynamics across all packing fractions while maintaining the diffusivity ratio $D_r=D_{\rm cold}/D_{\rm hot} \in \{10^{-4},1\}$. The time evolution of $i$-th particle is governed by Brownian dynamics as
\begin{align}
    \dot{\vec r}_i=\mu\left(\sum_{\substack{j =1 \\ j\neq i}}^N\vec{F}_{ij}+\vec{F}_{{\rm ext}}\right)+\vec{\eta}_i(t)\,,
\end{align}
where $\vec{r}_i \equiv (x_i,y_i)$ is the position of the $i$-th particle and  the mobility is considered  as $\mu=1$ for both types of particles. The variable $\vec{\eta}_i(t)$ is the Gaussian white noise with $\langle\eta_i^{\alpha}(t)\rangle=0$ and $\langle\eta_i^{\alpha}(t)\eta_j^{\beta}(t')\rangle=2D_i\delta_{ij}\delta_{\alpha \beta}\delta(t-t')$, where $\alpha,\beta$ corresponds to the Cartesian components and $D_i$ corresponds to the diffusivity of the $i$-th particle, which can be  $D_{\text{cold}}$ or $D_{\text{hot}}$ depending on the type of particle. The pairwise interaction force $ \vec{F}_{ij}=-\vec{\nabla}U(r_{ij})$ 
is derived from a short-range repulsive harmonic potential given by~\cite{McCarthy2024}
\begin{equation}
    U(r_{ij})=
    \begin{cases}
        0.5K_h (2\sigma-r_{ij})^2, & ~~~~r_{ij} < 2\sigma \\
        0, & ~~~~r_{ij} \geq 2\sigma
    \end{cases}
\end{equation}
where $r_{ij}=|\vec{r}_i-\vec{r}_j|$ measures the distance between the centers of the particles, the stiffness constant $K_h=50.0$, and $\sigma =1 $ is the radius of the particle, which also sets a length scale. 
The entire system is subjected to an externally applied potential
\begin{align}
    V_{\rm ext}(x)=V_0\cos(k_xx)\,,
\end{align}
which generates a force along the $x$-direction given by,  $F_{\rm ext}(x)=- dV_{\rm ext}(x)/dx = V_0 k_x\sin(k_xx)$.
Choosing $k_x=2\pi/L$ provides a single minima at the center of the box, i.e., at $x =L/2$ and $V_0$ sets the amplitude of the potential. $V_{\text{ext}}=V_0$ at the boundaries of the box. When $V_0=0$, there is no external potential and therefore no external force on the dynamics of the particles. The stochastic equations of motion are integrated using a second-order Runge-Kutta scheme~\cite{Braka1999} with a timestep $\Delta t=0.001$ in units of $\tau_0=\sigma^2/D_{\text{hot}}$. Further details of the update steps during simulation are provided in Appendix~\ref {label:AppA}. 
\begin{figure*}[t] 
    \centering
    \includegraphics[width=\textwidth]{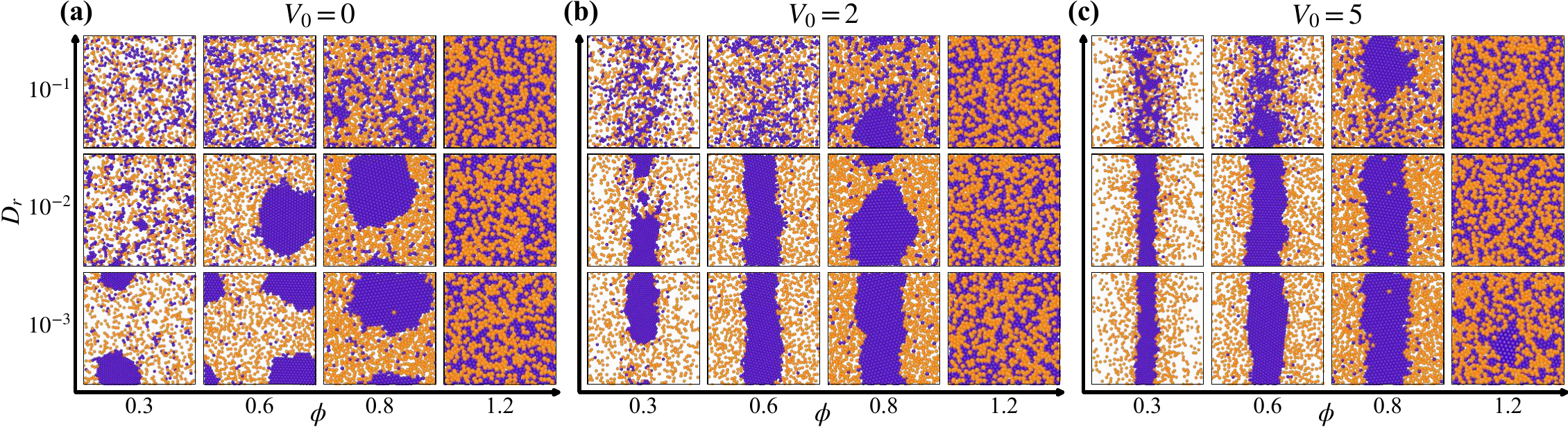}
    \caption{Typical steady-state snapshots of the binary system with particles having two different diffusivities (blue marking the ``cold'' and red marking the ``hot'' particles) for different packing fractions $\phi$ (along each column) and diffusivity ratios $D_r=D_{\text{cold}}/D_{\text{hot}}$ (along each row) are shown.  Different sets correspond to different choices of the external potential strength $V_0$, with (a) $V_0=0$, (b) $V_0=2$, and (c) $V_0=5$. At high $\phi=1.2$, for all values of $D_r$, the system does not show any segregation and remains in a mixed phase.}
    \label{fig:snapshots_ss}
\end{figure*}
 
We set $N=1000$, and $L$ is adjusted accordingly for the desired packing fraction $\phi=N\pi \sigma^2/L^2$. We vary packing fractions in the range $\phi \in [0.1,1.2]$.  However, due to the overlap of the particles,  the effective packing fraction can be lower than the initial value. All our results correspond to the case after the system reaches its steady state.  

%%%%%%%%%%%%%%%%%%%%%%%%%%%%%%%%%%%%%%%%%%%%%%%%%%%%%%%%%%%%%%%%%%%%%%%%%%%%%%%%%%%%%%%%%%%%%%%%%%%%%%%%%%%%%%%%%%%%%%%%%%%%%%%%%%%%
\section{Results}
%%%%%%%%%%%%%%%%%%%%%%%%%%%%%%%%%%%%%%%%%%%%%%%%%%%%%%%%%%%%%%%%%%%%%%%%%%%%%%%%%%%%%%%%%%%%%%%%%%%%%%%%%%%%%%%%%%%%%%%%%%%%%%%%%%%%
 
First, we investigate how variations in key parameters ($D_r, V_0, \phi$) affect the morphology of the system, specifically whether it remains in a mixed phase or it goes to a state where demixing or phase separation occurs between the two types of particles. In Fig.~\ref{fig:snapshots_ss}, we present typical snapshots of the steady-state configurations. For $V_0=0$, it is evident that the demixing occurs only when $D_r$ is sufficiently small. In particular, for $D_r=10^{-1}$, no demixing is observed for any packing fraction $\phi$. In contrast, for $D_r=10^{-2}$, the demixed phase emerges at $\phi \approx 0.6$. As $D_r$ decreases further, the onset of demixing shifts to a lower value of $\phi$. For example, for $D_r=10^{-3}$, demixed phase appears at $\phi \approx 0.3$. Interestingly, at a very high packing fraction, the system remains in a homogeneous or mixed phase regardless of the value of $D_r$ (see snapshots for $\phi=1.2$ in Fig.~\ref{fig:snapshots_ss}(a)). These findings are consistent with previous works~\cite{weber2016demix,kumari2017demixing}. Unlike in a granular system with inelastically colliding particles, where a percolating cluster arises due to dissipation of energy~\cite{paul2014dynamics,paul2017ballistic}, here the high-density cluster of ``cold'' particles forms only due to the asymmetric depletion forces, which are created due to diffusivity difference. These results provide a qualitative overview of the ranges of $\phi$ over which demixing occurs for different values of $D_r$. Furthermore, we find that cluster formation does not occur at any preferred location; instead, the cluster remains approximately symmetric and continuously migrates throughout the simulation box.

We now examine the effect of the external potential on the demixing behavior. As illustrated in Fig.~\ref{fig:snapshots_ss}(b-c), the complex interplay between particle dynamics and external potential gives rise to highly intriguing demixing behavior. Not only does it lead to demixing at lower values of $\phi$, the cluster tends to form a percolating band along the transverse direction of the applied potential, localized near the minima of the potential.

Notably, even for $D_r=10^{-1}$, for which no demixing is observed for $V_0=0$, the presence of the external potential induces phase separation at quite moderate packing fractions. Thus, the applied potential favors demixing and broadens the range of $D_r$ and $\phi$ over which the system demixes. For $V_0=5$, a percolating band of ``cold'' particles emerges even at $\phi=0.3$, localized near the potential minimum, in contrast to the diffusively moving cluster observed in the absence of any applied potential. The corresponding density profiles (presented in Appendix~\ref{append_densityprof}) for both ``hot'' and ``cold'' particles confirm this localization. Furthermore, for fixed $D_r$, we find that the stability of the cluster increases with increasing value of $V_0$.

%%%%%%%%%%%%%%%%%%%%%%%%%%%%%%%%%
\begin{figure*}[t]
    \centering
    \includegraphics[width=0.95\textwidth]{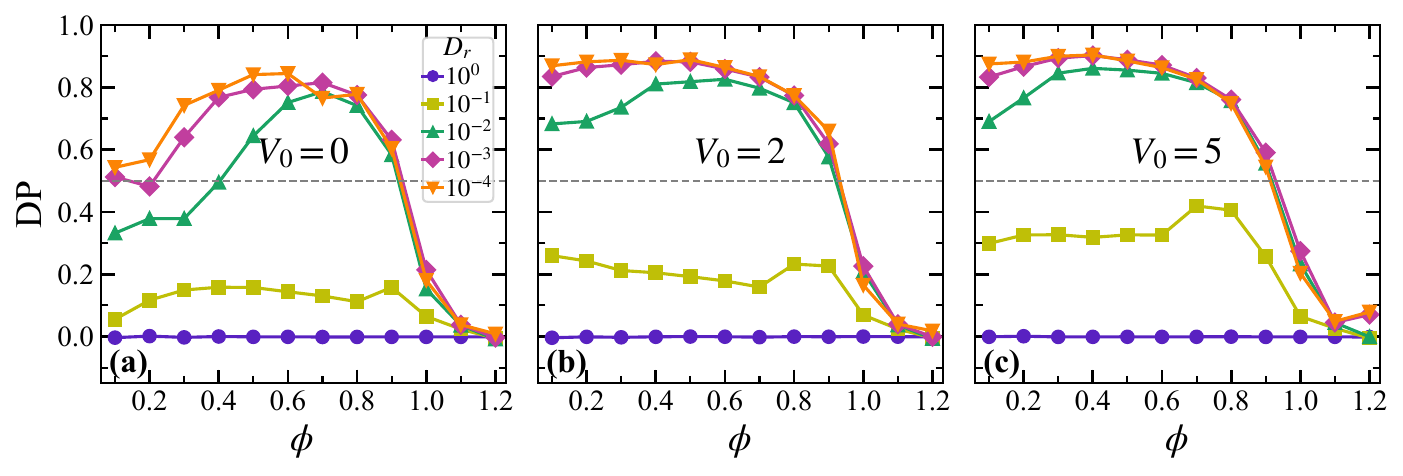}
    \caption{Demixing parameter (DP) of the binary mixture system for different values of packing fraction $\phi$ and diffusivity ratios $D_r$ for $V_0=0$ in (a) and in the presence of a potential, with $V_0=2$ in (b) and $V_0=5$ in (c). 
    $D_r=1$ corresponds to a mixed phase with DP $\approx 0$. Non-monotonic variation of DP versus $\phi$ shows a re-entrant mixing behavior of the system at large values of $\phi$.}
    \label{fig:ss_DP}
\end{figure*}

%%%%%%%%%%%%%%%%%%%%%%%%%%%%%%%%
\begin{figure}[htbp]
\centering
\includegraphics[width=0.95\linewidth]{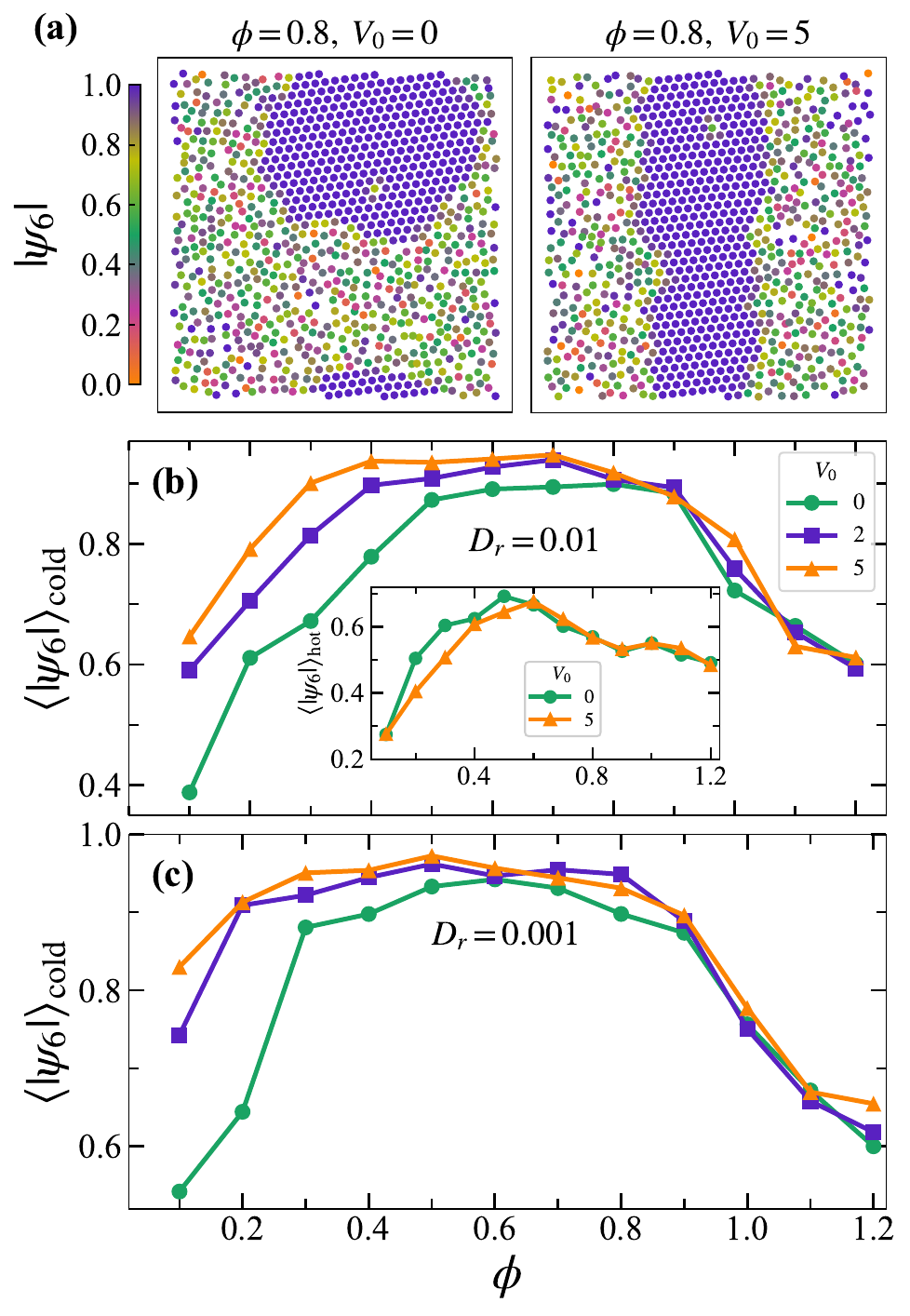}
\caption{(a) Representative snapshots marking the hexatic order corresponding to each particle of the system. The color bar indicates the hexatic parameter $|\psi_6|$ ranging between $[0,1]$. The presented snapshots are for $\phi=0.8$ with $D_r=10^{-3}$ for $V_0=0$ and $5$. Plots of the $\langle|\psi_6|\rangle_{\text{cold}}$ for ``cold'' particles versus $\phi$ for $D_r=0.01$ in (b) and for $D_r=0.001$ in (c) for three different values of $V_0$. $\langle|\psi_6|\rangle_{\text{cold}} \approx 0.9$ corresponds to nice hexagonal order as seen in (a). Inset in (b) shows $\langle|\psi_6|\rangle_{\text{hot}}$ for ``hot'' particles versus $\phi$ for $D_r=10^{-2}$ for $V_0=0$ and $5$. In both cases, value of $|\psi_6|$ is close to $\approx 0.6$. }
\label{fig:hexatic}
\end{figure}
%%%%%%%%%%%%%%%%%%%%%%%%%%

To quantify the degree of clustering, we calculate the demixing parameter (DP) defined as, 
\begin{equation}\label{demix_defn}
    \text{DP} = \langle \text{DP}_i \rangle = \Big\langle 2\Big(\frac{N_s^i}{N_t^i}-\frac{1}{2}\Big) \Big\rangle,
\end{equation}
where $\langle \cdots \rangle$ calculates the average over all particles in the system and over steady-state configurations. For any $i$-th particle, $N_t^i$ and $N_s^i$ count the total number of its neighbors and the number of neighbors of its same type, respectively. $\text{DP}=1$ represents a completely demixed state, in which $N_t^{i} \approx N_s^{i}$. However, if $N_t^{i} =2N_s^{i}$ , then \text{DP} $= 0$. Note that, according to Eq.~\eqref{demix_defn}, values of $\text{DP}$ slightly lower than $0$ is also possible. In Fig.~\ref{fig:ss_DP}(a)-(c) we plot DP versus $\phi$ for a few different values of $D_r$ for all our considered values of $V_0=0$, $2$ and $5$. For $D_r=1.0$, i.e., with $D_{\text{cold}}=D_{\text{hot}}$, we do not observe any demixing and DP remains close to $0$ throughout the range of $\phi$. For $V_0=0$ only for intermediate values of $\phi$, with $0.4 \le \phi \le 0.9$ the system shows demixing with the corresponding values of DP close to $\sim 0.8$. As $D_r$ decreases, the degree of demixing improves, and the value of DP increases.  However, for $\phi>0.9$, regardless of $D_r$, the system remains in a mixed phase. Now, for $V_0 > 0$, the demixing not only starts at a lower value of $\phi$ but the value of DP also reaches a value $\approx 0.9$, higher than that for $V_0=0$. However, for $\phi>1.0$ it is not possible to reach the demixed state, even if $V_0$ is very high. In fact, this occurs due to the availability of limited space; both the ``hot" and the ``cold" particles become trapped by their neighbors, and the ``cold" particles are unable to form any cluster or band structure. 

%%%%%%%%%%%%%%%%%%%%%%%%%%%%%
\begin{figure*}[t]
    \centering
    \includegraphics[width=0.95\textwidth]{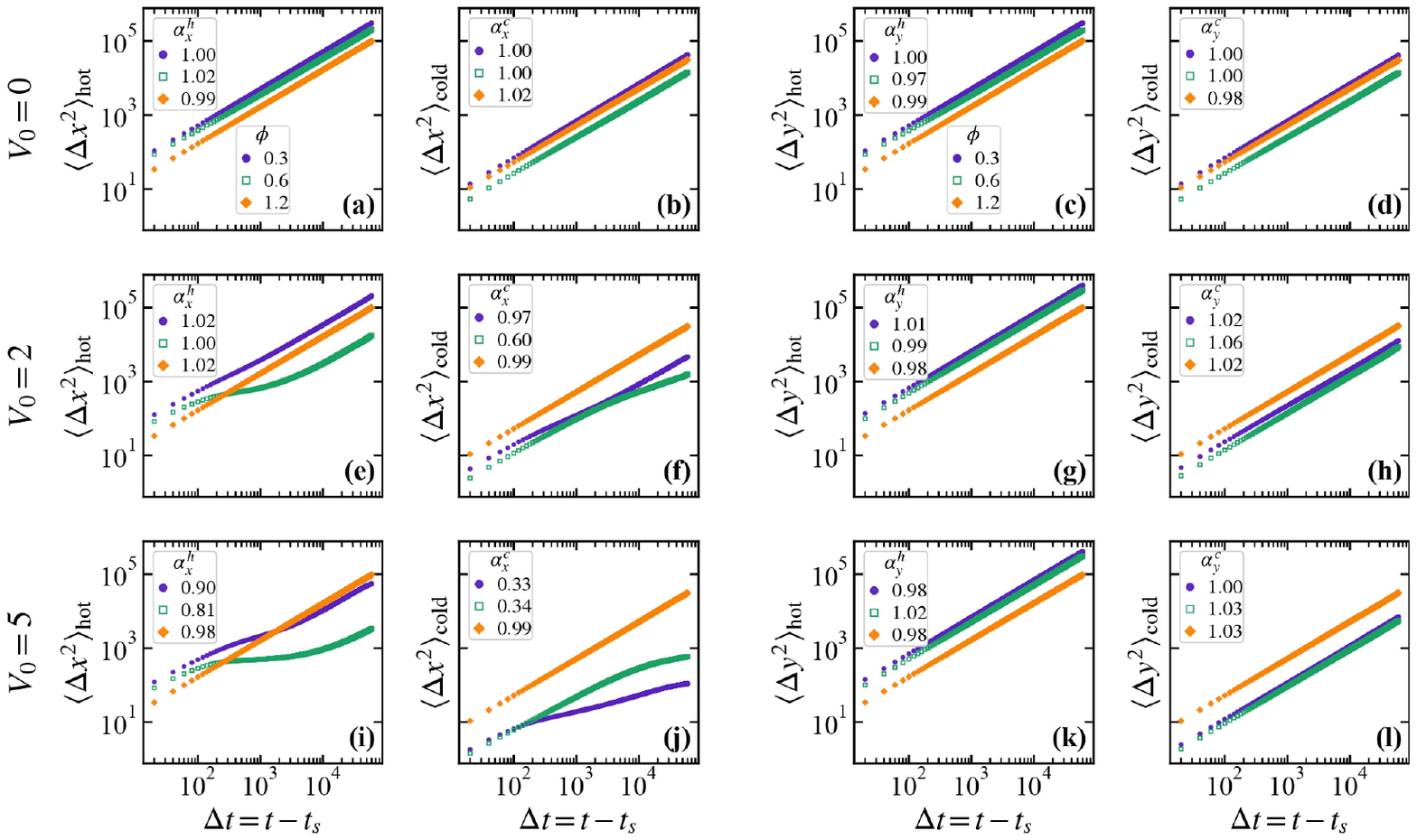}
    \caption{Mean-squared-displacement (MSD) of ``hot'' and ``cold'' tagged particle corresponding to $x$ and $y$ directions, i.e., $\langle \Delta x^2 \rangle_{\text{hot/cold}}$ and $\langle \Delta y^2 \rangle_{\text{hot/cold}}$ versus $\Delta t= (t-t_s)$ for different values of the packing fractions $\phi$ ($0.3, 0.6$ and $1.2$), for $V_0=0$ in (a)-(d), $V_0=2$ in (e)-(h) and $V_0=5$ in (i)-(l), as mentioned.  All data are for $D_r=0.01$ with $N=1000$. The corresponding values of the exponents $\alpha_x^{h,c}$ and $\alpha_y^{h,c}$ from the fitting of the late time data are mentioned in Table~\ref{tbl:msd_exponent_table}. }
    \label{fig:xy_msd_hotcold}
\end{figure*} 
%%%%%%%%%%%%%%%%%%%%%%%%%%%%%%
Now we provide a detailed quantitative structural arrangement of cold and hot particles by calculating the hexatic parameter $\psi_{6,i}$, which is defined as,
\begin{equation}
    \psi_{6,j} = \frac{1}{n_j} \sum_{k=1}^{n_j} e^{i 6 \theta_{jk}}\,,
    \label{eq:psi6}
\end{equation}
where $\theta_{jk}$ is related to the angle of the bond vector connecting the neighboring particles $j$ and $k$. The term $n_j$ defines the total number of local neighbors of $j$. The details of the calculation are mentioned in  Appendix~\ref{append_hexa}. 
$|\psi_{6,j}|=1$  indicates a perfect hexatically ordered crystal-like structure with quasi-long-range order. On the other hand, particles in the fluid phase will have, on average, $|\psi_{6,j}|\approx 0 $.

In Fig.~\ref{fig:hexatic}(a), we have shown representative snapshots marking the hexatic order parameter $|\psi_6|$ for all particles in the system. Snapshots are shown for $\phi=0.8$ and $D_r=10^{-3}$ with $V_0=0$ and $5$. The color bar represents the value of $|\psi_6|$ that varies from $0$ to $1$. It is evident from Fig~\ref{fig:hexatic}(a) that there is an almost perfect hexagonal closed-pack structure, $|\psi_6| \approx 1$, for the cold particles present within the cluster for both cases. In contrast to cold particles, the hot particles located outside the cluster exhibit higher mobility and do not exhibit the local structural arrangement (see the right panel of Fig.~\ref{fig:hexatic}(a)). In Fig.~\ref{fig:hexatic}(b-c), we show the variation of $\langle |\psi_6|\rangle_{\text{cold}}$ for the ``cold'' particles  as a function of $\phi$ for different $D_r$ and three different values of $V_0$. $\langle \cdots \rangle$ corresponds to the average over all ``cold'' particles. As shown in Fig.~\ref{fig:hexatic}(b-c), for $V_0=0$ and in a very low regime of $\phi$, the system remains in a mixed phase; however, an external potential significantly induces demixing and improves $\langle |\psi_6|\rangle_{\text{cold}} $. For $V_0=2$ and $V_0=5$, this increase is more than 1.5 times the value observed at $V_0=0$ (when $D_r=0.001$ and $\phi=0.2$); see Fig.~\ref{fig:hexatic}(c). However, for a large enough $\phi$, the system does not rely on $V_0$ and remains in a mixed state. For moderate to larger $\phi$ ($0.7 \le \phi \le 1.0$), the system shows demixing with $\langle |\psi_6|\rangle_{\text{cold}} \approx 0.9$, confirming the hexagonal arrangement of the particles.  Furthermore, Figs~\ref{fig:snapshots_ss}  and ~\ref{fig:ss_DP} confirm that, for any value of $D_r$, the value of $\phi$ at which demixing begins decreases with $V_0$.
 As seen from the snapshots as well as from Fig.~\ref{fig:ss_DP}, for any value of $D_r$, the value of $\phi$ at which demixing begins, decreases with $V_0$. 
 A similar hexagonal arrangement can also be confirmed using a pair correlation function or a structure factor, which show discrete peaks corresponding to the fixed distances between the particles \cite{dolai2018phase}. 
 In the inset of Fig.~\ref{fig:hexatic}(b), we show  $\langle |\psi_6|\rangle_{\text{hot}}$ corresponding to the ``hot'' particles for $V_0=0$ and $5$. For any value of $\phi$, a low value of $\langle |\psi_6|\rangle_{\text{hot}}$ indicates a disordered arrangement of hot particles.
 %%%%%%%%%%%%%%%%%%%%%%%%%
\begin{figure}[htbp]
    \centering
    \includegraphics[width=0.48\textwidth]{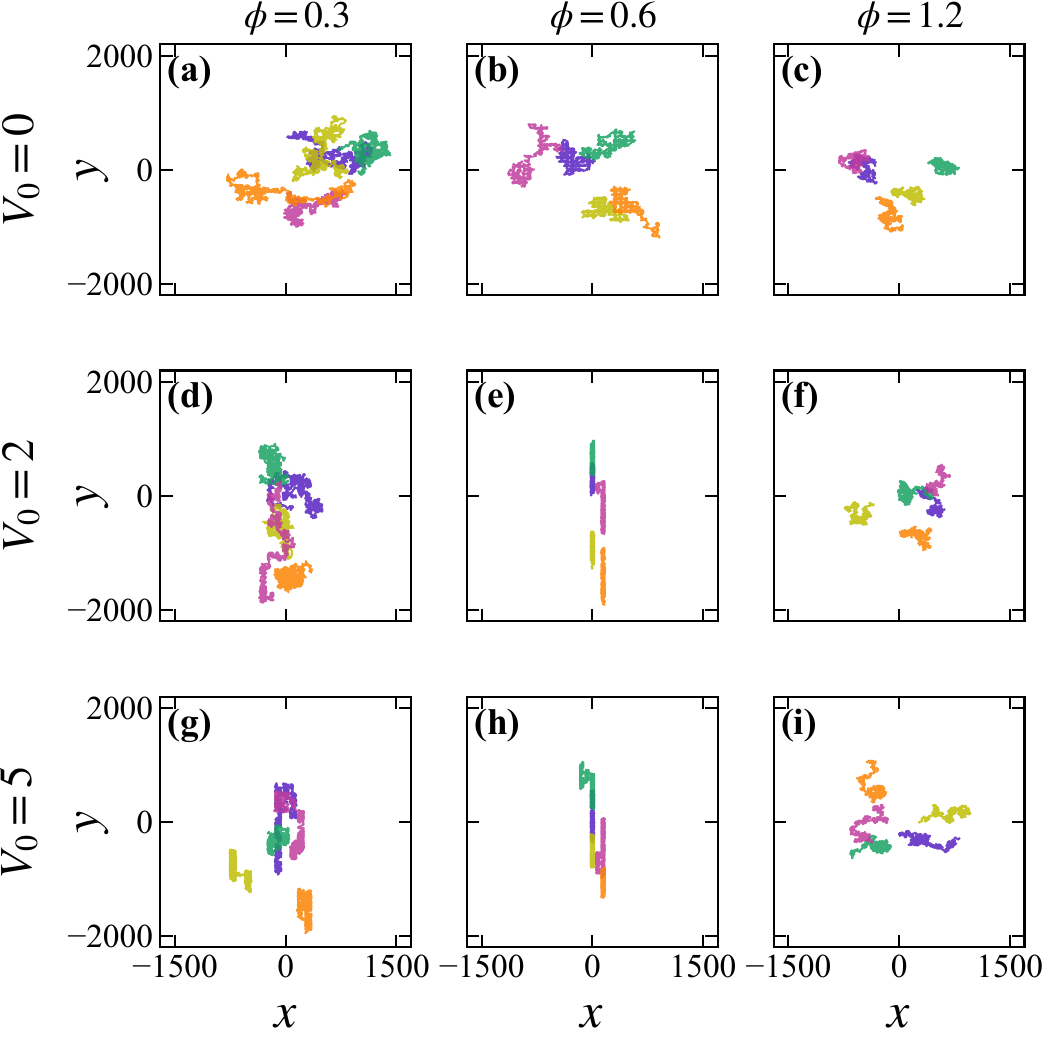}
    \caption{Typical trajectories of a few ``hot'' particles after the system reaches its steady-state for different values of the packing fraction $\phi=0.3, 0.6,1.2$. (a)-(c) corresponds to $V_0=0$, (d)-(f) are for $V_0=2$ and (g)-(i) are for $V_0=5$.  }
    \label{fig:hot_xy_trajec}
\end{figure} 
%%%%%%%%%%%%%%%%%%%%%%%%

\begin{table*}[t]
\renewcommand{\arraystretch}{1.5}
\setlength{\tabcolsep}{20pt}
\caption{Scaling exponents $\alpha$ for the MSD for ``hot'' and ``cold'' particles along $x$ and $y$ directions.}
\label{tbl:msd_exponent_table}

\begin{tabular*}{\linewidth}{@{\extracolsep{\fill}}c c c c c c}
\hline\hline
Packing Fraction ($\phi$) & Potential ($V_0$) & $\alpha_x^h$ & $\alpha_y^h$ & $\alpha_x^c$ & $\alpha_y^c$ \\
\hline\hline

0.3 & 0 & 0.9961 & 1.0008 & 1.0004 & 1.0003 \\
%\hline
0.3 & 2 & 1.0205 & 1.0122 & 0.9747 & 1.0172 \\
%\hline
0.3 & 5 & 0.8994 & 0.9793 & 0.3252 & 0.9978 \\
\hline
0.6 & 0 & 1.0219 & 0.9668 & 1.0018 & 1.0004 \\
%\hline
0.6 & 2 & 0.9969 & 0.9890 & 0.6027 & 1.0607 \\
%\hline
0.6 & 5 & 0.8091 & 1.0164 & 0.3448 & 1.0302 \\
\hline
1.2 & 0 & 0.9889 & 0.9924 & 1.0163 & 0.9797 \\
%\hline 
1.2 & 2 & 1.0160 & 0.9774 & 0.9878 & 1.0221 \\
%\hline
1.2 & 5 & 0.9766 & 0.9795 & 0.9873 & 1.0305 \\
\hline\hline

\end{tabular*}
\end{table*}

In order to quantify dynamic behavior across various time-scales and how the motion of the particles get influenced due to the fluctuating complex environments in different phases, we have calculated MSD, defined as
\begin{equation}
\text{MSD}(t) = \Big \langle \big[\vec{r}(t_0 + t) - \vec{r}(t_0)\big]^2 \Big \rangle\,,
\end{equation}
where $\vec{r}(t)$ is the position of a particle at time $t$, and $\langle \dots \rangle$ denotes averaged over many starting configuration and over the
time origin $t_0$. In general, MSD follows a power-law behavior $\text{MSD} \sim t^{\alpha}$, where the exponent $\alpha$ determines the dynamics. While $\alpha=2$ indicates ballistic behavior, $\alpha=1$ represents normal diffusive dynamics. Super-diffusion occurs when $1 < \alpha < 2$. If the motion of the particle slows down, then $0 < \alpha < 1$, and the particle is known to exhibit sub-diffusive behavior. Signatures of super- and sub-diffusive motions of particles are typically observed in the presence of various interactions, i.e. in crowded environments \cite{sokolov_2012}.

As shown in Fig~\ref{fig:snapshots_ss} in the presence of an external potential $V_0$, the ``cold'' particles accumulate near the minima of the potential following the nucleation of the cluster and then grow in the $y$-direction. The width of the cluster along the $x$-direction depends on the packing fraction $\phi$. Furthermore, when the value of $D_r$ is sufficiently low, and the mobility of ``cold" particles decreases, the ``band" becomes percolating in the $y$-direction. The anisotropy in the growth of clusters in the system is reflected in the  MSD of ``hot'' and ``cold'' particles (see Fig.~\ref{fig:xy_msd_hotcold}). The first two columns of Fig.~\ref{fig:xy_msd_hotcold} represent the time evaluation of MSD of hot $\langle \Delta x^2 (t)\rangle _{\text{hot}}$ and cold  $\langle \Delta x^2 (t)\rangle _{\text{cold}}$ particles in the $x $ direction. The corresponding exponents are marked as $\alpha_x^h$ and $\alpha_x^c$.  Similar abbreviations are used for the $y$-direction as well. Here, $t_s$ being the starting time of measurement after the system reaches its steady state.

For $V_0 = 0$, the system is in a mixed state at $\phi = 0.3$ and $\phi = 1.2$, whereas at $\phi = 0.6$, the ``cold" particles form a cluster, and the system exhibits a demixed state.  As shown in Figs.\ref{fig:xy_msd_hotcold}(a)-(d), regardless of the values of $\phi$, the magnitude of the MSD of ``hot'' particles in any direction is always greater than the corresponding values of the ``cold'' particles; this is consistent with the fact that ``hot" particles exhibit higher diffusivity. Moreover, the magnitude of MSD  for “hot” particles in any direction decreases with increasing $\phi$. However, for ``cold'' particles, a non-monotonic behavior with $\phi$ is observed in the amplitude of MSD. For $\phi=0.6$ when the ``cold'' particles are within the cluster, the corresponding amplitudes of  $\langle \Delta x^2 \rangle _{\text{cold}}$ and $\langle \Delta y^2 \rangle _{\text{cold}}$ appear lower compared to the other values considered of $\phi$ for which the system remains homogeneous.

In the presence of an external potential $V(x)$, the symmetry in the system is broken (see Fig.~\ref{fig:snapshots_ss} and Fig.~\ref{fig:density_profl} in Appendix~\ref{append_densityprof}). As seen, the transverse direction of the applied potential is preferred for the ``cold'' particles to accumulate and grow near the center of the box. Unlike the case of Vicsek-like no-size particles \cite{Vicsek1995,gaur2026global}, here the soft-core repulsion among particles does not allow them to overlap, and thus  the cluster also expands in the $x$-direction, keeping the center of the band near the minima. As illustrated in the second and third rows of Fig.~\ref{fig:xy_msd_hotcold}, there is a significant difference in the dynamics of the ``hot and ``cold" particles, which is attributed to the formation of a stable cluster.

The ``hot'' particles remain outside the band and in a dilute phase. As evident in Fig.~\ref{fig:xy_msd_hotcold}, the late-time behavior shows diffusion in both directions; the intermediate regime exhibits different dynamical features. $\langle \Delta x^2 \rangle_{\text{hot}}$ shows an intermediate plateau similar to the caging behavior observed in glassy systems or very dense active systems \cite{berthier2019glassy}. This is due to the non-motile ``cold''  cluster working as a reflecting layer for the ``hot'' particles. Even though the motion of ``hot'' particles gets restricted along the $x$-direction, diffusive motion along the  $y$-direction is not affected much. However, in the case of $V_0=2$,  with $\phi=0.3$ and $1.2$, MSD for any direction does not show any anomalous behavior. The corresponding plots for $\langle \Delta r^2(t) \rangle= \langle \Delta x^2(t) \rangle + \langle \Delta y^2(t) \rangle$ versus $\Delta t$ for different values of $V_0$ and $\phi$ are shown in  Appendix.~\ref{label:AppA}. $\langle \Delta r^2 (t)\rangle_{\text{hot/cold}}$ do not show any anisotropy and anomalous behavior. 

Next, we examine the dynamics of ``cold" particles for $V_0=2$ and $5$. Due to confinement along the $x$-direction, $\langle \Delta x^2 \rangle_{\text{cold}}$ within the band shows a sub-diffusive motion. However, along the $y$-direction, in which the cluster spans, $\langle \Delta y^2 \rangle_{\text{cold}}$ shows a diffusive motion (see Fig.~\ref{fig:xy_msd_hotcold}(f,j).)
In Figs~\ref{fig:xy_msd_hotcold}(e-j), $\langle \Delta x^2 \rangle_{\text{hot}}$ shows a larger plateau and $\langle \Delta x^2 \rangle_{\text{cold}}$ shows a stronger sub-diffusive behavior. For instance, if $\phi =0.6$  and $V_0=5$ the MSD exponent of the cold particle $\alpha_x^c \approx 0.34$ is significantly lower ( about $50 \%$) that of $V_0=5$. Furthermore, for $\phi =0.3$  and $V_0=5$, $\alpha_x^c \approx 0.33$ is one-third when compared to $V_0=2$. For $\phi = 0.3$, the small value of $\alpha_x^c$ follows directly from the fact that, due to the presence of the strong potential $V_0=5$, a fully separated and percolating band is formed.

Now, we focus our attention on the nature of the MSD, as illustrated in Fig.~\ref{fig:xy_msd_hotcold}(j). The lower value of $\langle\Delta x^2\rangle_\text{cold}$ for $\phi = 0.3$ compared to $\phi = 0.6$ indicates that the bandwidth becomes more prominent relative to the packing fraction. For $\phi=0.3$, due to the narrow band, ``cold" particles become trapped more densely. However, when $\phi=0.6$, the bandwidth widens, and the particles have more space, resulting in a higher value of $\langle\Delta x^2\rangle_\text{cold}$. Furthermore, the applied potential is unidirectional and exerts no effective influence in the transverse direction; consequently, the magnitudes of $\langle \Delta y^2 \rangle_{\text{hot}}$ and $\langle \Delta y^2 \rangle_{\text{cold}}$ show the diffusive nature, see Fig.~\ref{fig:xy_msd_hotcold}. The corresponding exponents $\alpha_h^x$, $\alpha_h^y$, for ``hot'' particles and $\alpha_c^x$, $\alpha_c^y$ for ``cold'' particles, obtained from the fitting of various MSD data at late times, are listed in Table~\ref{tbl:msd_exponent_table}.

As observed, the MSD data of tagged particles provide intriguing insights into the dynamic anisotropy of ``hot" and ``cold" particles in the presence of a potential. The aforementioned plateau and sub-diffusive motion can be better understood by examining the trajectories of the ``hot" particles and the cluster's overall dynamics. Regarding this, we first show in Fig.~\ref{fig:hot_xy_trajec} the typical trajectories of $5$ tagged ``hot'' particles for different values of $\phi$ and $V_0$. For $V_0=0$, for any value of $\phi$, when the particles perform diffusive Brownian dynamics, their trajectories look ``uniform'' in both directions. Also, the typical ranges covered along $x$ and $y$ directions decrease as $\phi$ increases.  However, in the presence of potential, non-uniform behavior observed along $x$ and $y$ directions, see Fig.~\ref{fig:hot_xy_trajec}(d)-(i). For $V_0=5$, with $\phi=0.3$ and $0.6$, when the motion of the particles becomes confined in the $x$-direction, resulting in the formation of an intermediate plateau, the temporal span depends upon the width and the boundary fluctuations of the ``cold'' band. This fact is quite visible from the trajectories shown in Fig.~\ref{fig:hot_xy_trajec}(g)-(h). The fluctuations in the positions of the boundary particles depend on the strength of the potential. This also explains that the lower amplitude of $\langle \Delta x^2 \rangle_{\text{hot}}$ for $\phi=0.6$ than that for $\phi=0.3$ as observed from Fig.~\ref{fig:xy_msd_hotcold}(i). This fact is also reflected from Figs.~\ref{fig:xy_msd_hotcold}(j) and (l), as for these, $\langle \Delta x^2\rangle_{\text{cold}}$ shows a sub-diffusion in comparison to a normal diffusion for $\langle y^2\rangle_{\text{cold}}$ for $\phi=0.3$ and $0.6$. The dynamics of particles near the band edge also depends on the stability of the band-forming particles and on fluctuations in their positions.

%%%%%%%%%%%%%%%%%%%%%%%%%%%%%%%%%%%%%%%%%%%%%
\begin{figure}[htbp]
\centering
\includegraphics[width=0.95\linewidth]{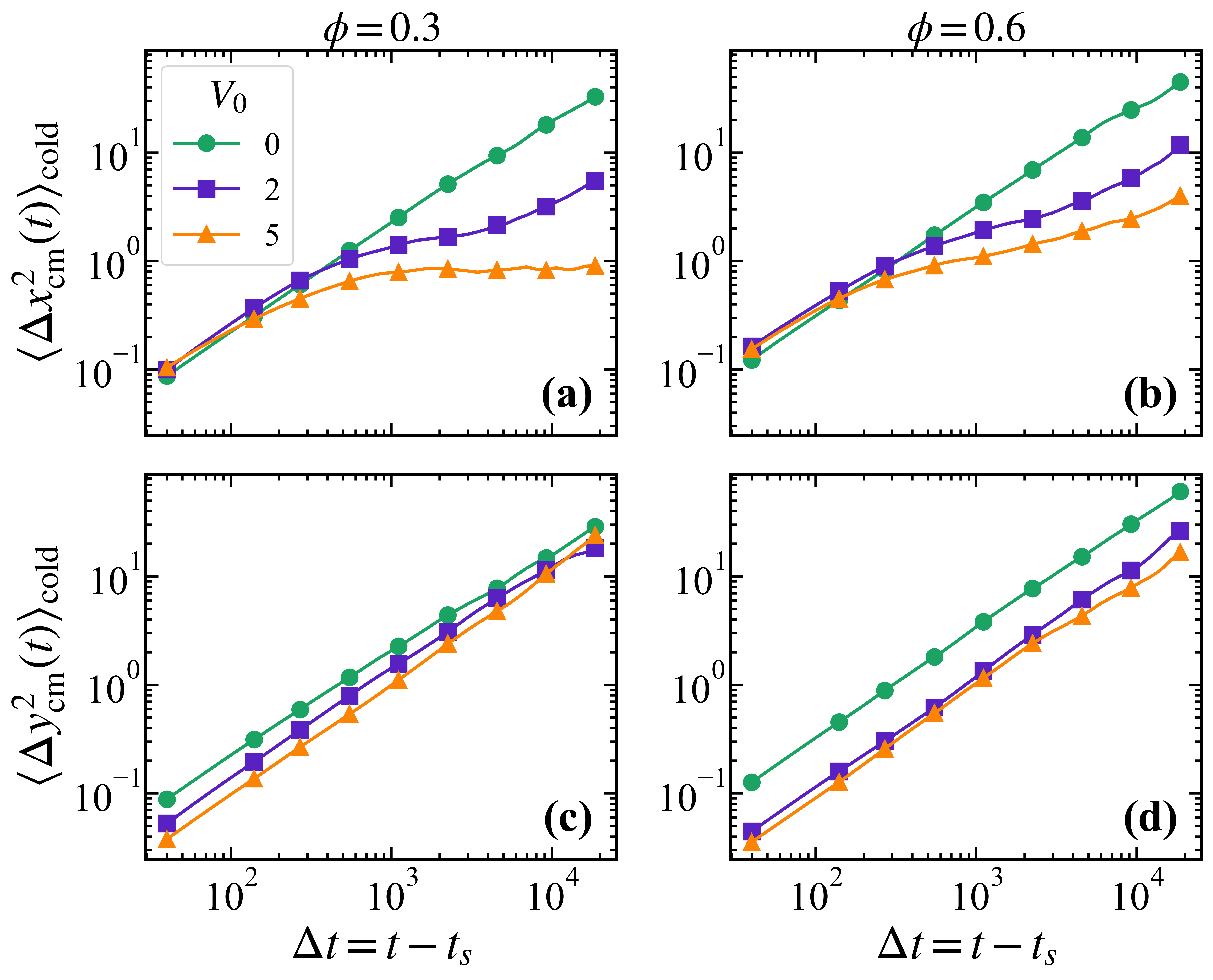}
\caption{MSDs $\langle \Delta x_{\text{cm}}^2(t) \rangle _{\text{cold}}$ and $\langle \Delta y_{\text{cm}}^2(t) \rangle _{\text{cold}}$ corresponding to $x$ and $y$-directions of the centre-of-mass of the cluster made of ``cold'' particles versus $\Delta t=t-t_s$ for different values of $V_0$ for $\phi=0.3$ and $\phi=0.6$, as mentioned. All the presented data are for $D_r=10^{-2}$.}
\label{fig:msd_xy_coldcm}
\end{figure}

% ===================== FIGURE 1 =====================
\begin{figure*}[!t]
\centering

\includegraphics[width=0.49\textwidth,page=1]{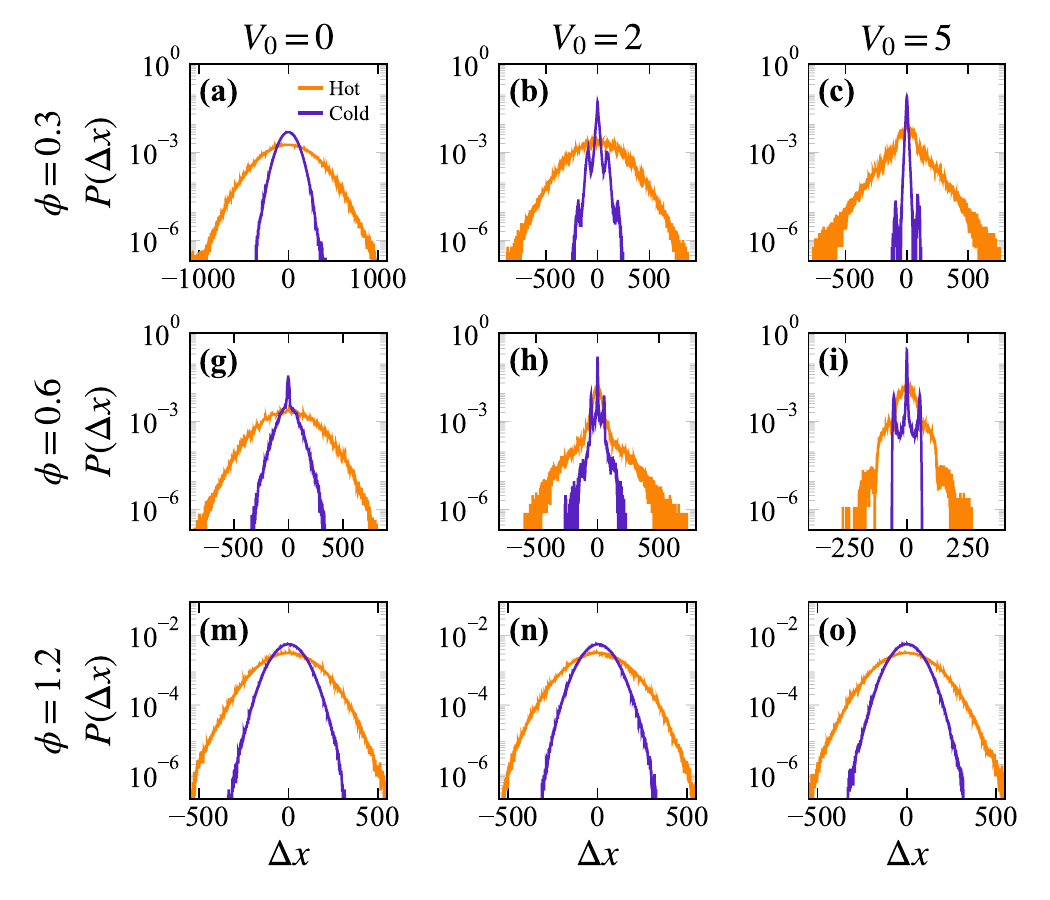}
\hspace{0.001\textwidth}
\includegraphics[width=0.49\textwidth,page=2]{Fig7.pdf}
\caption{(a) Normalized distributions of $P(\Delta x)$ versus the displacement along $x$-direction $\Delta x$ for different values of $\phi$ for $V_0=0,2$ and $5$. In each frame data are shown for both ``hot'' and ``cold'' particles. Choices of $\phi$ are consistent with the snapshots in Fig.~\ref{fig:snapshots_ss} as well as data for Fig.~\ref{fig:xy_msd_hotcold}. (b) Similar to (a), but for the displacement $\Delta y$ along the $y$-direction.  All our data are shown for $D_r=0.01$, for the displacements calculated over a time separation $\Delta t=10^4$. All the presented data are with $D_r=0.01$ and the distributions are averaged over $N_c=N_h=500$ particles.}
\label{fig:distri_delx_dely_tagged}
\end{figure*}

%=========================================================

The dynamics of the ``cold'' cluster can also provide better insight on the dynamical anisotropy observed for ``hot'' and ``cold'' particles. 
The centre-of-mass (cm) of the cluster formed by ``cold'' particles is defined as: $\vec{r}_{\text{cm}}^{\text{cold}}= \sum_{i=1}^{N_c'} \Vec{r}_i /N_c'$, where $N_c'$ corresponds to the number of ``cold'' particles which participate in the formation of the cluster. In Fig.~\ref{fig:msd_xy_coldcm}  we show MSD of the centre-of-mass of the cluster corresponding to $x$ and $y$-direction, i.e.,   $\langle \Delta x^2_{\text{cm}}(t)\rangle_{\text{cold}}$  and $\langle \Delta y^2_{\text{cm}}(t)\rangle_{\text{cold}}$ versus $t-t_s$, respectively, for different values of $V_0$. Using a local density cut-off,  the cluster corresponding to ``cold'' particles is identified by discarding the ``hot'' particles, as well as a few ``cold'' particles that are in the vapor phase. Data for MSD of the cm are shown for packing fractions $\phi=0.3$ and $0.6$ for different values of $V_0$ for $D_r=0.01$, see Fig.~\ref{fig:msd_xy_coldcm}. The corresponding trajectories $\vec{r}_{\text{cm}}^{\text{cold}} \equiv (x_{\text{cm}}^{\text{cold}}, y_{\text{cm}}^{\text{cold}})$ of the cluster are shown in Fig.~\ref{fig:com_cold_traj} in Appendix~\ref{append_traj_coldcm}. In these trajectories, the asymmetry in motion in different directions is quite evident. These non-uniform behavior also appear in the plots  of $\langle \Delta x^2_{\text{cm}}(t)\rangle_{\text{cold}}$ and $\langle \Delta y^2_{\text{cm}}(t)\rangle_{\text{cold}}$. Data for both the MSDs show diffusive behavior for $V_0=0$ for both values of $\phi$, whether in a mixed or de-mixed phase. Following the motion of the cluster, the MSD of the ``cold'' particles for both directions always shows diffusive motion, as seen from Fig.~\ref{fig:xy_msd_hotcold}. In the presence of potential,  whereas data corresponding to  $\langle \Delta y^2_{\text{cm}}(t)\rangle_{\text{cold}}$ for both values of $\phi$ show diffusive motion (see (c) and (d)),  anomalous sub-diffusive dynamics is observed for $\langle \Delta x^2_{\text{cm}}(t)\rangle_{\text{cold}}$. The degree of sub-diffusion increases with the strength of the applied potential (as seen from (a) and (b)).  Sub-diffusion of $\langle \Delta x^2_{\text{cm}}(t)\rangle_{\text{cold}}$ was also hinted from the MSD of individual tagged particles (see Fig.~\ref{fig:xy_msd_hotcold}(j)) as the motion along $x$-direction is restricted within the width of the band.  Also, in the demixed phase, for any $\phi$, the amplitude of $\langle \Delta x^2_{\text{cm}}(t)\rangle_{\text{cold}}$  decreases with increasing $V_0$, i.e.,  in presence of more stable cluster. On the other hand, the ``cold'' particles can move throughout the system along the $y$-direction as the band percolates. Thus, $\langle \Delta y^2_{\text{cm}}(t)\rangle_{\text{cold}}$ shows a diffusive motion even in the presence of the potential. However,  for any value of $\phi$, the amplitude of $\langle \Delta y^2_{\text{cm}}(t)\rangle_{\text{cold}}$ decreases with increasing $V_0$, which corresponds to a more stable band with fewer fluctuations of the particles. Similar anisotropic behavior can be more prominent with the lowering of the value of $D_r$.

Our discussions so far, related to MSD of tagged particles and the cluster or band, provide a good rationale for the dynamical anisotropy. However, the MSD is related only to the second moment of the displacement distributions. It is expected that there may be qualitative differences and non-Gaussian features in the distributions. To provide a clearer quantitative picture, we show normalized distributions ($P(\Delta x)$ and $P(\Delta y)$) of the displacements along the $x$ and $ y$ directions, respectively, in Fig.~\ref{fig:distri_delx_dely_tagged} for various values of $\phi$ and $V_0$. To facilitate comparison, the system parameters selected here are identical to those shown in Fig.~\ref{fig:xy_msd_hotcold}. Also, in each frame, distributions for ``hot'' and ``cold'' particles are shown together for better visualization. For all the data presented, $\Delta x$ and $\Delta y$ correspond to the displacements obtained for the time separation $\Delta t=t-t_s=10^4$ along its trajectory. Each distribution is averaged over all particles of that type and over independent $10$ runs.

It is worth mentioning here that, irrespective of the dynamics of the particles depending upon whether the system remains in a mixed or demixed phase, displacements of ``cold'' particles are always less compared to the ``hot'' ones. This can be easily understood from the respective ranges of $\Delta x$ or $\Delta y$, see Fig.\ref{fig:distri_delx_dely_tagged}. For $V_0=0$, irrespective of whether the system is in a mixed ($\phi=0.3$ and $1.2$) or in a demixed ($\phi=0.6$) state, the distributions $P(\Delta x)$ and $P(\Delta y)$ for both ``hot'' particle distributions always appear as Gaussian. Fitting with Gaussian functions is shown in Fig.~\ref{fig:distri_delx_dely_gaussfit} in  Appendix~\ref{append_gaussin_fit}. However, for the ``cold'' particles, the distributions slightly change depending upon whether the steady-state configuration is homogeneous or phase-separated. Compared to a fully Gaussian distribution for the homogeneous or mixed phase, $P(\Delta x)$ and $P(\Delta y)$ show a higher peak value near $\Delta x, \Delta y \approx 0$ as seen from the data for $\phi=0.6$ (see Figs.\ref{fig:distri_delx_dely_tagged}~(g)-(l)). And also,  as seen from the distributions,  the ranges of $\Delta x$ and $\Delta y$ decrease with increasing value of $\phi$. 

Now in the presence of $V_0$ also, whereas the late time behavior of $P(\Delta y)$ for ``hot'' particles is always Gaussian, the behavior of  $P(\Delta y)$ for ``cold'' particles changes. In Fig.~\ref{fig:distri_delx_dely_tagged}, $P(\Delta y)$ ``cold'' particles shows a much higher probability near $0$ for $V_0=2,5$. These are more or less similar to the case $P(\Delta x)$ for ``cold'' particles forming a spontaneous cluster for $\phi=0.6$ with $V_0=0$. Note that, in case of $V_0 \ne 0$,  while the cluster of ``cold'' particles becomes band-like with finite width along the $x$-direction and percolating along $y$-direction, the displacements along $x$-direction of both ``hot'' and ``cold'' particles get restricted, as discussed earlier. For the ``cold'' particles, along with a higher peak observed for $P(\Delta x)$ close to $0$, the other two smaller peaks appear near the boundaries of the band (see Figs.\ref{fig:distri_delx_dely_tagged}~(h) and (i) for $V_0=2$ and $5$, respectively). By increasing $V_0$, the boundary fluctuations of the band decrease. A steeper boundary near the minima of the potential acts somewhat like a finite support for the less-mobile ``cold'' particles. The effect of the boundary is also manifested in the trajectories of the ``hot'' particles as their motion gets restricted along $x$-direction, as seen from Fig.~\ref{fig:hot_xy_trajec}(h)-(i). In fact, similar behavior in $P(\Delta x)$ and $P(\Delta y)$ is also observed in (c), as $\phi=0.3$ with $V_0=5$ leads to demixing. For these, the distributions $P(\Delta x)$ for the ``hot'' particles become more stretched exponential-like compared to the Gaussian distribution for $P(\Delta y)$. Also, the ranges for $\Delta x$ are smaller than those for $\Delta y$ (see the plots corresponding to $\phi=0.3, V_0=5$ in (c) and (f) or $\phi=0.6, V_0=2$ in (h) and (k)). However, the distributions $P(\Delta x)$, $P(\Delta y)$  for both ``hot'' and ``cold'' particles with $\phi=1.2$ are always Gaussian, suggesting a homogeneous or mixed phase and diffusive nature of particles.  Also, in such higher value of $\phi$, it appears that the ranges of $\Delta x$ and $\Delta y$ do not depend upon $V_0$. Fitting with Gaussian forms are shown in Appendix~\ref{append_gaussin_fit}.

%%%%%%%%%%%%%%%%%%%%%%%%%%%%%%%%%%%%%%%%%%%%%%%%%%%%%%%%%%%%%%%%%%%%%%%%%
\section{Summary and Conclusion}
%%%%%%%%%%%%%%%%%%%%%%%%%%%%%%%%%%%%%%%%%%%%%%%%%%%%%%%%%%%%%%%%%%%%%%%%
Using Brownian dynamics simulations, we have studied the demixing properties of a binary system with differences in their diffusivities and interacting via soft-core repulsive potential. By systematically varying the ratio of their diffusivities $D_r$ and the overall packing fraction $\phi$ of the particles, the properties of the system and the dynamics of both high diffusive (``hot'') and less diffusive (``cold'') particles are analyzed in their corresponding steady-state in the presence of a spatially varying external potential acting along one direction. Whereas the phase-separation is measured by the density profile and the demixing parameter, the dynamics in different directions is probed by the corresponding mean-squared-displacement (MSD). In the absence of any external potential, ``cold'' particles form a dense cluster, solely due to the asymmetry in the depletion forces. However, the presence of an external potential creates an energy barrier that facilitates accumulation of particles near the minima, thereby promoting the formation of a band. Consequently, the system exhibits demixing behavior at a lower value of $\phi$ and a higher value of $D_r$ compared to the $V_0=0$ case. Furthermore, for sufficiently higher value of $\phi$, the band becomes percolating, spanning the system along transverse direction of the applied potential.

It is obvious that the dynamics of the ``cold'' particles get restricted by the finite width of the band formed near the potential minima and show a sub-diffusive motion along the $x$-direction compared to its long-time free diffusion along the $y$-direction, i.e., transverse direction of the applied potential. Thus, the ``cold'' particles show anisotropy in their dynamics in different directions. Interestingly, even though the ``hot'' particles are always in the vapor phase, the presence of such a band affects their dynamics as well. Whereas their motion along $y$-direction is diffusive, motion along the $x$-direction gets restricted due to the stable band and thus shows an intermediate plateau in their corresponding MSD before following a late-time diffusive motion. These facts have also been confirmed from the MSD of the center-of-mass of the cluster in each direction as well as from their Gaussian versus non-Gaussian position fluctuations. In the presence of a stable ``immobile'' band, even though the ``hot'' particles feel the band as a potential barrier or a reflecting wall, the band works as a finite support to the ``cold'' particles along  $x$-direction. 

 As mentioned, our main aim was to study the nature of dynamical anisotropy and its role in leading to non-Gaussian fluctuations. However, relating this dynamical anisotropy to the percolation probability is another important consideration \cite{chiara_bc24_pre_fluidflow}, as the demixing behavior and clustering  strongly depend on the mobility difference and the strength of the potential barrier.  Understanding or controlling these can yield interesting effects on transport in the system. In this work, we have examined the dynamics in the steady state, i.e., after the system reaches its demixed or mixed state. However, it can be interesting to see the mechanism and dynamics of cluster formation. Even though, for $V_0=0$, the cluster resembles a MIPS one, as observed in self-propelled particles, here, in the presence of only diffusion, the diffusive coalescence mechanism can still be dominant. As observed, the particles form a nice hexagonal ordering within the cluster. During such a nonequilibrium process, equal-time and two-time correlations can be investigated to understand the emergence of particle correlations. Moreover, velocity ordering associated with micro-flocking behavior has been reported in several related systems \cite{levis2018micro,claudio20_mipsmicro,paul2024spontaneous}. It would therefore be interesting to investigate how the presence of a confining potential influences the emergence of micro-flocking behavior within the “cold” cluster.

 Overall, our proposed mechanism provides a generic route for controlling diffusivity-driven phase separation and designing soft materials with tunable segregation properties, with potential applicability in biomolecular condensates and optoelectronic systems where external fields can be used to manipulate structure and organization.

%%%%%%%%%%%%%%%%%%%%%%%%%%%%%%%%%%%%%%%%%%%%%%%%%%%%%%%%%%%%%%%%%%%%%%%%%%%%%%%%%%%%%%%%%
\section{Acknowledgment}
RT acknowledges the fellowship provided by the Ministry of Education (MoE), Government of India. SP acknowledges University of Delhi for providing financial assistance through the Faculty Research Programme under Grant-IOE (Ref.\ No. IOE/2024-25/12/FRP). SK acknowledges the financial support received from the Anusandhan National Research Foundation, India ( EEQ/2023/000676) and IIT Jodhpur for a research initiation grant (I/RIG/SNT/20240068). 
%%%%%%%%%%%%%%%%%%%%%%%%%%%%%%%%%%%%%%%%%%%%%%%%%%%%%%%%%%%%%%%%%%%%%%%%%

\section{Conflicts of Interest}
There is no conflicts of interest to declare.

\section{Data and codes Availability}
Data and codes can be available upon request to the authors.

\appendix
% ============================================================
\section{ Numerical Integration Scheme}\label{label:AppA}

\renewcommand{\theequation}{A\arabic{equation}}
\setcounter{equation}{0}

Particles evolve according to overdamped Brownian dynamics, where thermal fluctuations are incorporated through Gaussian random displacements. For each particle, the stochastic noise follows $\langle \eta_i^{\alpha}(t)\rangle =0$ and $\langle \eta_i^{\alpha}(t) \eta_j^{\beta}(t')\rangle=2D_i \delta_{ij}\delta_{\alpha\beta} \delta(t-t')$, i.e., each component $\eta_i^{\alpha}$ follows a Gaussian random variable with zero mean and unit variance.   
A second-order Runge-Kutta scheme~\cite{Braka1999} is employed to integrate the equations of motion. $\vec{r}_i(t+\Delta t)$ from the known $\vec{r}_i(t)$.  First we estimate an intermediate position at time \( t + \Delta t \),
\begin{equation}
\vec{r}_i^{\,\prime} = \vec{r}_i(t)
+ \mu\left[\vec{F}_i(t) + \vec{F}_i^{\mathrm{ext}}(t) \right]\Delta t
+ \vec{\eta}_i
\end{equation}
where \(\vec F_i(t)\) denotes the total force arising from pairwise interactions.  Using the intermediate positions \(\vec r_i^{\,\prime}\), the interaction and external forces at time \(t+\Delta t\) are recalculated, and the final particle positions are obtained as
\begin{equation}
\begin{aligned}
\vec{r}_i(t+\Delta t) = \vec{r}_i(t)
&+ \frac{\mu}{2} \Big[
\vec{F}_i(t) + \vec{F}_i(t+\Delta t)\Big]\Delta t\\
&+ \frac{\mu}{2} \Big[\vec{F}_i^{\mathrm{ext}}(t)
+ \vec{F}_i^{\mathrm{ext}}(t+\Delta t)
\Big]\Delta t + \vec{\eta}_i
\end{aligned}
\end{equation}
Simulations are performed until the system reaches a steady state. The total force acting on each particle includes contributions from interparticle interactions and the external spatially periodic potential described in the main text.

\FloatBarrier
\section{Method of estimating the neighbors and hexatic order parameter }\label{append_hexa}

We take ``hot" and ``cold" particles separately and use the cKDTree method~\cite{Bentley1975} to find neighboring particles within a cutoff distance of $r_c=2.5\sigma$ for each particle. For a given particle $i$, all neighboring particles $j$ are identified, and the bond angle $\theta_{ij}$ between the vector connecting particles $i$ and $j$ and a fixed reference axis is calculated. These angles are mapped into complex numbers using $ e^{i6\theta_{jk}} $ where the factor 6 is due to the sixfold rotational symmetry of a hexagonal lattice. If all neighbors are arranged in a perfect hexagonal pattern, all these complex numbers align, and their average has a magnitude close to $1$.
If the system is disordered, its value will be $0$; we consider only its magnitude, as the phase angle is irrelevant here. We take the average of all the particles of each type, the corresponding average numbers represent the degree of local hexagonal ordering of ``hot'' and ``cold'' particles separately.
%%%%%%%%%%%%%%%%%%%%%%%%%%%%%%%%%%%%%%%%%%%%%%%%%%%%%55
\FloatBarrier
\section{ Density Profiles of ``hot'' and ``cold'' Particles in Presence of Potential}\label{append_densityprof}
\begin{figure}[htbp]
    \centering
    \includegraphics[width=1\columnwidth]{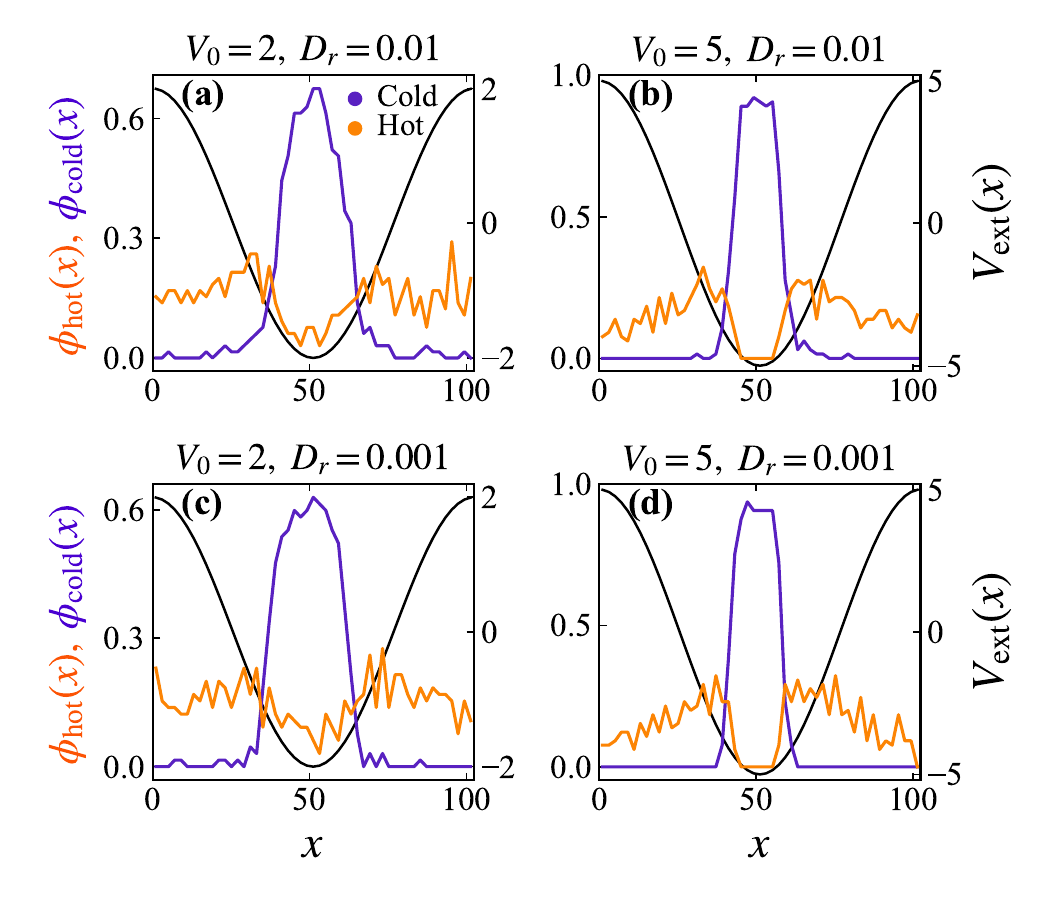}
    \caption{Spatial arrangment of hot ($\phi_{\text{hot}}(x)$) and cold ($\phi_{\text{cold}}(x)$) particles as a function $x$-direction in presence  of $V_{\text{ext}(x)} =V_0 \sin (k_xx)$ with $V_0=2$ and $5$. (a)-(b) corresponds to data with $D_r=0.01$ and  (c)-(d) are with $D_r=0.001$. The black lines in each plot show the variation of $V_{\text{ext}}(x)$. $\phi_{\text{cold}}(x)$ shows peak near $V_{\text{ext}}=0$, i.e., at $x \approx L/2$. $\phi_{\text{hot}}(x)$ always remains low with a deep valley near the potential minima. A higher potential strength corresponds to a more stable band with fewer fluctuations in its boundaries.}
    \label{fig:density_profl}
\end{figure}

To compute the spatial distribution of particles, we take the simulation box, which is divided into  a number of vertical bins with a fixed width, that is 
$\Delta x = 2\sigma$. The total number of bins $ N_{\text{bins}} = \frac{L}{\Delta x}$. 
The number of particles is counted within the area of each bin 
$ A_{\text{bin}} = L \times \Delta x $ and each particle contributes an area $\phi(x) = \frac{N_{\text{bin}} \cdot \pi a^2} { L \cdot \Delta x}$. As shown in  Fig.~\ref{fig:density_profl}, for a fixed value of $D_r$, as the potential increases a sharp and thin band of ``cold" particles is formed. This sharp peak indicates that, with increasing potential, the majority of the ``cold" particles drift towards the minima and tend to form a compact band-like structure. However, for a given value of $V_0$, reducing $D_r$ facilitates the formation of a stable cluster, leading to a wider band of ``cold" particles.

%%%%%%%%%%%%%%%%%%%%%%%%%%%%%%

%%%%%%%%%%%%%%%%%%%%%%%%%%%%%%%%%%%%%%%%%%%%%%%%%%%%%
%%%% Appendix D
%%%%%%%%%%%%%%%%%%%%%%%%%%%%%%%%%%%%%%%%%%%%%%%%%%%%%%%%%

\section{MSD $\langle \Delta r^2(t)\rangle$ for ``cold'' and ``hot'' particles for different values of $\phi$ and $V_0$}\label{append_fullmsd}

\begin{figure}[htbp]
\centering
\includegraphics[width=0.45\textwidth]{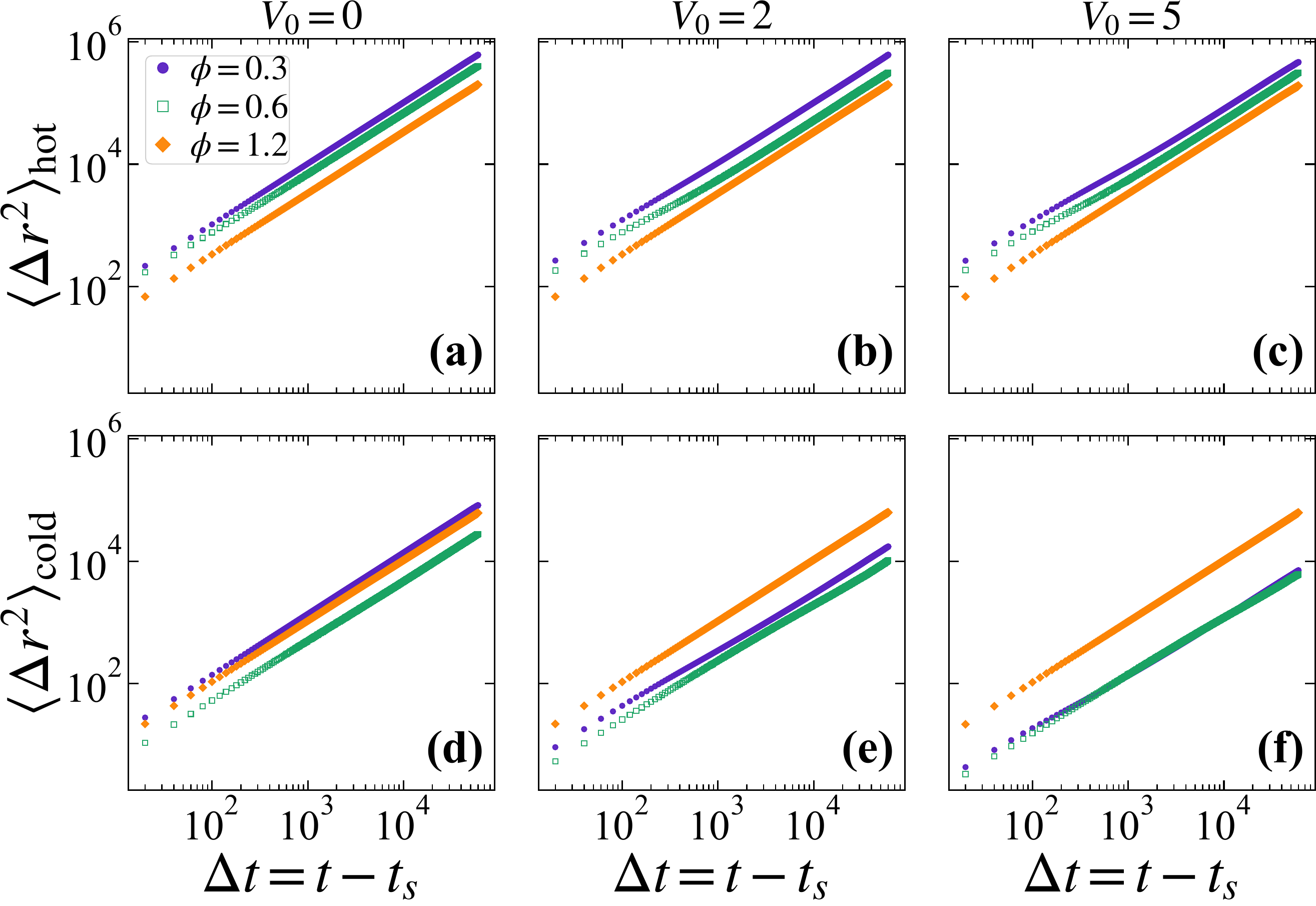}
\caption{Mean-squared displacement $\langle \Delta r^2(t)\rangle$ versus $\Delta t=t-t_s$ of the ``hot'' and ``cold'' particles for different values of $V_0$. In each sub-plot data are shown for three different values of $\phi$ for comparison.  The corresponding individual plots of MSD along $x$ and $y$ directions are presented in Fig.~\ref{fig:xy_msd_hotcold}.}
\label{fig:msd_full_hotcold}
\end{figure}

As seen from Fig.~\ref{fig:xy_msd_hotcold} the MSD of ``hot'' and ``cold'' particles for different directions exhibit different behavior, indicating dynamical anisotropy in the steady-state dynamics in the presence of a potential. The full MSD is defined as  $\langle \Delta r^2(t)\rangle=\langle \Delta x^2(t)\rangle+\langle \Delta y^2(t)\rangle $. In Fig.~\ref{fig:msd_full_hotcold} we plot $\langle \Delta r^2(t)\rangle$  and $\langle \Delta r^2(t)\rangle$ versus $\Delta t=t-t_s$ for different values of the packing fraction $\phi$ and $V_0$. Both ``hot'' and ``cold'' particles exhibit diffusive behavior with their corresponding exponents close to $1$ and MSD for ``hot'' particles for all values of $V_0$ have higher amplitude than those for ``cold'' particles.  Also, for hot particles, amplitude of $\langle \Delta r^2(t)\rangle_{\text{hot}}$ decreases with increasing $\phi$, as the particles are always in ``vapor'' phase. Dynamical anisotropy observed for ``hot'' particles in different directions due to band formation and in the presence of a potential can not be readily accounted from the full $\langle \Delta r^2(t)$. Even though the MSD for ``cold" particles exhibits diffusive behavior, its magnitude depends on $\phi$ i.e., whether the system is in a mixed or a demixed state. Lower amplitudes of $\langle \Delta r^2(t)\rangle_{\text{cold}}$ for $\phi=0.3$ or $0.6$ in the presence of potential $V_0$ indicate the ``cold'' particles are within the band. 

%%%%%%%%%%%%%%%%%%%%%%%%%%%%%%%%%%%%%%%%%%%%%%
\FloatBarrier
% ============================================================
\section{Trajectories of the center-of-mass of the ``cold'' particle cluster}\label{append_traj_coldcm}
\begin{figure}[htbp]
\centering
\includegraphics[width=\linewidth]{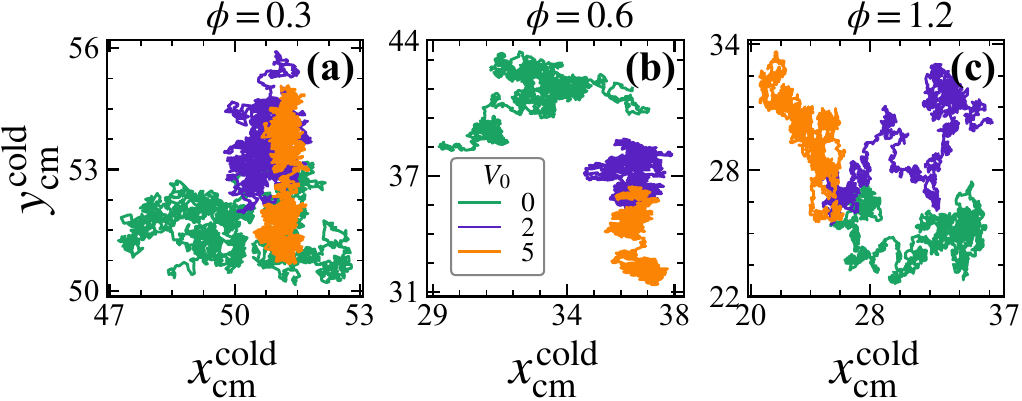}
\caption{
Representative trajectories of the center-of-mass (cm), calculated for the ``cold'' particles, for different values of $V_0$. (a)-(c) correspond to different values of $\phi$. These trajectories help in visualizing the anisotropies in motion of as quantified via $\langle \Delta x^2_{\text{cm}}(t)\rangle_{\text{cold}}$  and $\langle \Delta y^2_{\text{cm}}(t)\rangle_{\text{cold}}$ in Fig.~\ref{fig:msd_xy_coldcm}.}
\label{fig:com_cold_traj}
\end{figure}

Similarly to MSD plots shown in Fig.~\ref{fig:msd_xy_coldcm}, we calculated the trajectories relative to the center-of-mass (cm) of the ``cold'' cluster. In Fig.~\ref{fig:com_cold_traj}(a)-(c) we show the trajectories corresponding to the cm as $x_{\text{cm}}$ versus $y_{\text{cm}}$ for different values of $\phi$. 

%%%%%%%%%%%%%%%%%%%%%%%%%%%%%%%%%%%%%%%%%%%%%%%%%%%%%%%%%%%%%%%%

For $\phi=0.3$ with $V_0=0$ the cm shows diffusive behavior as there is no cluster formation. For $V_0=2$ and $5$, as shown in Figure~\ref{fig:msd_xy_coldcm} and corresponding to the sub-diffusion and diffusion represented by $\langle \Delta x^2_{\text{cm}}(t)\rangle_{\text{cold}}$ and $\langle \Delta y^2_{\text{cm}}(t)\rangle_{\text{cold}}$, we observe that the trajectories are more directed and spread out in the $y$-direction compared to the $x$-direction. This can also be observed from the range of $x$ and $y$-directions. Similar behavior is also observed for $\phi=0.6$. However, for $\phi=1.2$ when the particle is in a homogeneous state, the trajectories always show Brownian-like behavior.

% ============================================================
\FloatBarrier
\numberwithin{equation}{section}
\renewcommand{\theequation}{\thesection\arabic{equation}}

\begin{figure}[t!]
\centering
\includegraphics[width=0.45\textwidth,page=1]{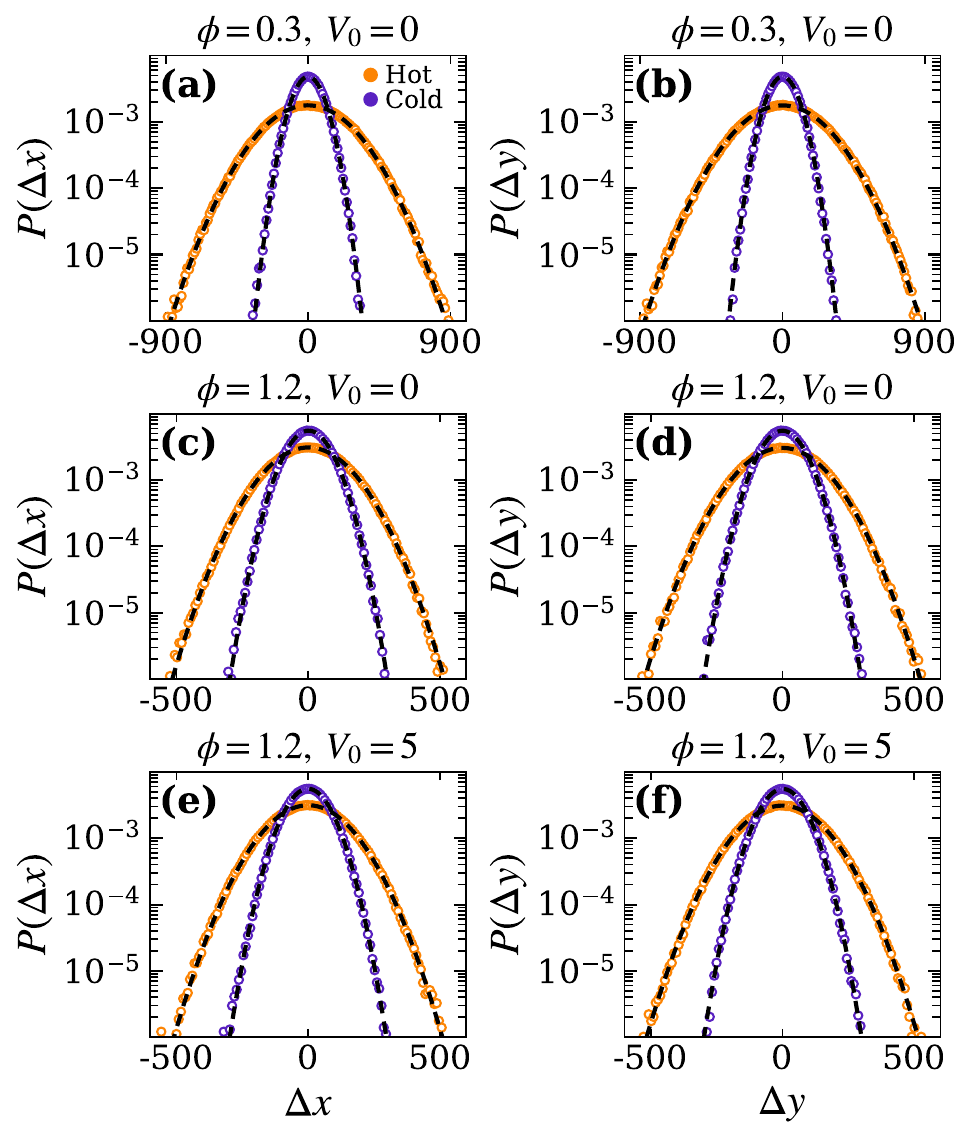}
\caption{Normalized distributions $P(\Delta x)$ and $P(\Delta y)$ correspond to the $x$ and $y$ directions, for $V_0=0$ with $\phi=0.3$ and $1.2$ in (a)-(d), and for $V_0=5$ with $\phi=1.2$ in (e)-(f). Each panel consists of data for both ``hot'' and ``cold'' particles. The dashed lines correspond to Gaussian fitting data as listed in Table~\ref{tb:tbl2}. }
\label{fig:distri_delx_dely_gaussfit}
\end{figure}

%%%%%%%%%%%%%%%%%%%%%%%%%%%%%%%%
\setcounter{section}{5}
\section{ Fitting with Gaussian forms}\label{append_gaussin_fit}
The trajectory of the particles is quantified from their probability distributions, which are presented in Fig.~\ref{fig:distri_delx_dely_tagged}. In the case of diffusion, the corresponding distributions are expected to be Gaussian. Here, particularly for diffusive cases, we present the distributions and fit them with Gaussian functions. Data are shown for $V_0=0$ with $\phi=0.3$ and $\phi=1.2$, for which the system remains in the mixed state. Even in the presence of a potential $V_0=5$, for $\phi=1.2$ the particles show diffusion. The distributions are fitted with ~\cite{Lemaitre2023} and shown in Fig.~\ref{fig:distri_delx_dely_gaussfit}
\begin{equation}
P(m)=\frac{1}{\sqrt{2\pi \sigma_m^2}}
\exp\left(-\frac{(m-\mu_m)^2}{2\sigma_m^2}\right)
\end{equation}
where variable $m$ corresponds to $\Delta x$ or $\Delta y$ for the ``hot'' or ``cold'' particles.  $\mu_m$ and $\sigma_m$ represent the mean and standard deviation of the corresponding distributions, respectively. The values of corresponding $\bar{x}$ and $\sigma$ are mentioned in Table~\ref{tb:tbl2}. To fit this function to the numerically obtained distribution data, the non-linear least-squares fitting method is used. The initial Gaussian is taken directly from the data, and the fitting method iteratively adjusts the parameters to minimize the difference between the Gaussian and the distribution data.

%%%%%%%%%%%%%%%%%%%%%%%%%%%%%%%
\begin{table}

\caption{Gaussian fitting parameters for hot and cold particles at different packing fractions ($\phi$), external potentials ($V_0$), and directions.}
\label{tb:tbl2}
\begin{tabular}{ccccccc}
\hline
$\phi$ & $V_0$ & Direction & $\mu_{hot}$ & $\sigma_{hot}$ & $\mu_{cold}$ & $\sigma_{cold}$\\
\hline
0.3 & 0 & $x$ & -1.372 & 225.140 & -0.146 & 83.12\\
0.3 & 0 & $y$ & 0.692 & 225.798 & 0.345 & 83.02\\
1.2 & 0 & $x$ & 0.777 & 128.894 & 1.551 & 71.94\\
1.2 & 0 & $y$ & -0.669 & 130.721 & -0.074 & 72.10\\
1.2 & 5 & $x$ & -0.257 & 126.973 & -0.678 & 71.58\\
1.2 & 5 & $y$ & -0.565 & 128.554 & 0.721 & 72.02\\
\hline
\end{tabular}
\end{table}

%%%%%%%%%%%%%%%%%%%%%%%%%%%%%%%%%%%%%%%%%%%%%%%%%%%%%%%

\FloatBarrier
%

%\bibliography{PRA.bib} 

\end{document}